# Possible evidence for spin-transfer torque induced by spin-triplet supercurrent

Lailai Li[1,3], Yuelei Zhao[2]*, Xixiang Zhang[2], and Young Sun[1,3]*
[1]Beijing National Laboratory for Condensed Matter Physics, Institute of Physics, Chinese Academy of Sciences, Beijing 100190, People's Republic of China
[2]King Abdullah University of Science and Technology (KAUST), Thuwal 23955-6900, Saudi Arabia
[3]School of Physical Science, University of Chinese Academy of Sciences, Beijing 100190, People's Republic of China
*email: youngsun@iphy.ac.cn; yuelei.zhao@kaust.edu.sa

Cooper pairs in superconductors are normally spin singlet (spin antiparallel). Nevertheless, recent studies suggest that spin-triplet Cooper pairs (spin parallel) can be created at carefully engineered superconductor-ferromagnet interfaces[1-4]. If Cooper pairs are spin-polarized they would transport not only charge but also a net spin component, but without dissipation, and therefore minimize the heating effects associated with spintronic devices. Although it is now established that triplet supercurrents exist, their most interesting property – spin – is only inferred indirectly from transport measurements[5-10]. In conventional spintronics, it is well known that spin currents generate spin-transfer torques that alter magnetization dynamics and switch magnetic moments. The observation of similar effects due to spin-triplet supercurrents would not only confirm the net spin of triplet pairs but also pave the way for applications of superconducting spintronics. Here, we present a possible evidence for spin-transfer torques induced by triplet supercurrents in superconductor/ferromagnet/superconductor (S/F/S) Josephson junctions. Below the superconducting transition temperature $T_C$, the ferromagnetic resonance (FMR) field at X-band (~ 9.0 GHz) shifts rapidly to a lower field with decreasing temperature due to the spin-transfer torques induced by triplet supercurrents. In contrast, this phenomenon is absent in ferromagnet/superconductor (F/S) bilayers and superconductor/insulator/ferromagnet/superconductor (S/I/F/S) multilayers where no supercurrents pass through the ferromagnetic layer. These experimental observations are discussed with theoretical predictions for ferromagnetic Josephson junctions with precessing magnetization.

Spin-triplet supercurrents combining superconducting and magnetic orders provide a great opportunity to enhance the functionality and performance of spintronic devices by offering the possibility of long-range spin-polarized supercurrents. As spin-one triplet Cooper pairs, unlike singlet pairs, can carry a net spin component, a spin-polarized current is naturally associated with the triplet supercurrents. Meanwhile, spin-one triplet Cooper pairs are immune to pair breaking by the exchange field in ferromagnets so that triplet Cooper pairs sustain long-range correlations in spintronic devices. To use such triplet supercurrents in spintronics it is necessary to effectively generate and manipulate triplet pairs in devices.

In the past decade, a number of theoretical models have been proposed to explain how spin-polarized supercurrents can be created and controlled in S/F heterostructures[11-20], with key ingredients ranging from non-uniform superconductor, inhomogeneous and non-collinear magnetization to strong spin-orbital coupling, *etc.* The first experimental evidence for long-range triplet supercurrents was reported by Keizer *et al.*[5] from the observation of supercurrent passing through the half-metallic ferromagnet $CrO_2$. Then a series of experiments demonstrated systematic evidence for triplet supercurrents in S/F/S Josephson junctions[6-10]. Although these existing experiments provide compelling evidences for triplet pairing in S/F heterostructures, they are not directly probing or using the spin carried by triplet supercurrents.

A well-known and useful phenomenon in spintronics is the spin-transfer torque induced by spin-polarized currents, which has been widely used to switch magnetization and control magnetization dynamics[21]. Similarly, the triplet supercurrents are anticipated to induce spin-transfer torques when passing through a ferromagnet. The demonstration of spin-transfer torques due to triplet supercurrents would not only confirm the net spin of triplet pairs but also pave the way for the emergence of superconducting spintronics. Recently, there are quite a number of theoretical works addressing on the spin-transfer torques and magnetization dynamics related to triplet supercurrents[22-31]. However, the experimental studies lie well behind theoretical progresses. So far no clear experimental evidence for spin-transfer torques induced by triplet supercurrents has been reported.

In this work, we have investigated magnetization dynamics in a series of S/F/S Josephson junctions via FMR technique. The results demonstrate a significant influence of superconductivity on magnetization dynamics: the resonance field ($H_r$) shifts rapidly to a lower field below superconducting transition temperature $T_c$. In contrast, such an effect is absent in the control experiments performed on S/F bilayers and S/I/F/S multilayers where no supercurrents pass through the device. Therefore, our experiments provide an evidence for the spin-transfer torques induced by triplet supercurrents.

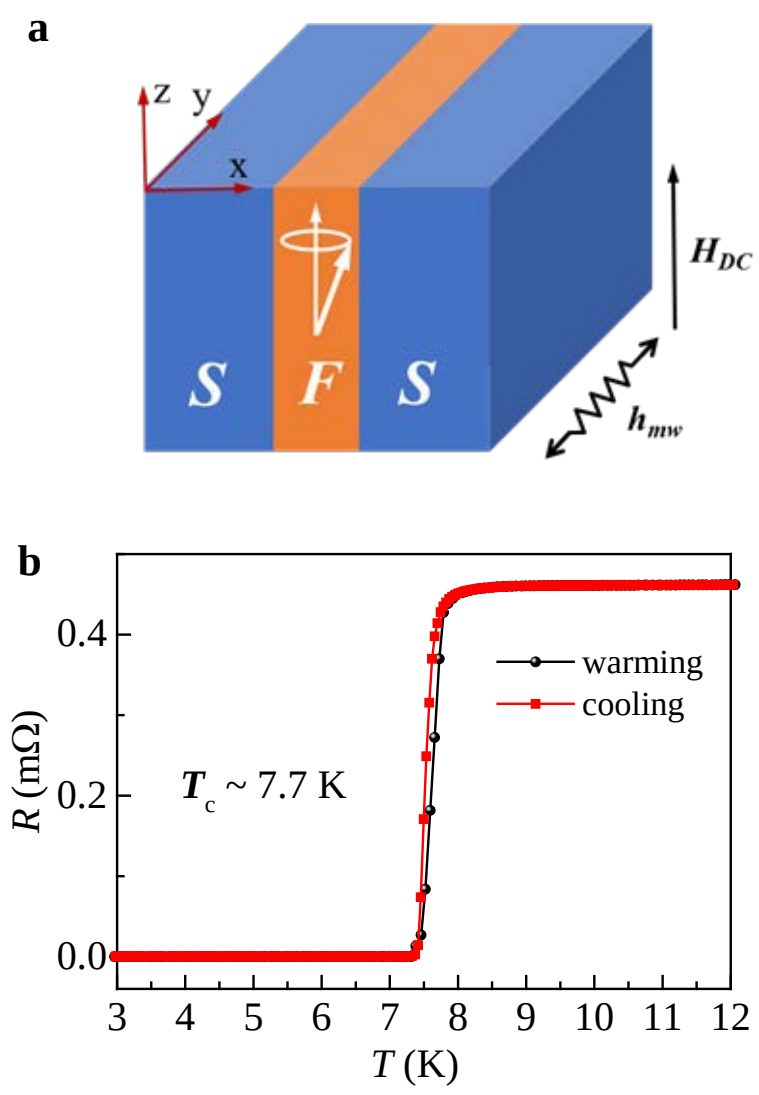

**Figure 1 Geometry of the ferromagnetic Josephson junctions and the configuration of FMR measurements. a,** The Josephson junctions in this study consist of a ferromagnetic layer ($Ni_{80}Fe_{20}$, 5-30 nm in thickness) and two superconductor layers (Nb, 100 nm in thickness). The FMR is measured at a fixed microwave frequency (~ 9 GHz) while scanning the DC magnetic field applied in the film plane. **b,** The resistance of a Nb film as a function of temperature. The superconducting transition temperature $T_C$ is ~ 7.7 K.

The geometry of the ferromagnetic Josephson junctions and the schematic of FMR are shown in Fig. 1a. When a DC magnetic field $\boldsymbol{H}_{DC}$ is applied not along the direction of magnetization, the magnetization will rotate to the direction of $\boldsymbol{H}_{DC}$ along spiral path by the driven torque and damping torque. If a microwave field with magnetic component $\boldsymbol{h}_{mw}$ perpendicular to $\boldsymbol{H}_{DC}$ is applied, the magnetization can absorb microwave energy and precess continuously in balance with the damping torque. This is the basic principle of FMR. In our study, the ferromagnetic Josephson junctions are made of two superconducting layers of Nb (100 nm in thickness) and a FM layer of $Ni_{80}Fe_{20}$ (5 - 30 nm in thickness). As shown in Fig. 1b, the transport measurement suggests a superconducting transition temperature $T_c$ ~ 7.7 K of the Nb layer.

The FMR spectra of a Nb(100 nm)/$Ni_{80}Fe_{20}$(20 nm)/Nb(100 nm) Josephson junction measured at X-band (~ 9 GHz) are shown in Fig. 2a. All the resonance lines exhibit a single Lorenz line-shape. The resonance field $H_r$ changes little with temperature above $T_c$. However, $H_r$ shifts rapidly to a lower field with decreasing temperature below $T_c$. The temperature dependence of $H_r$ is plotted in Fig. 2b. As temperature decreases from $T_C$ (7.7 K) to 4.2 K, $H_r$ shifts by ~ 70 mT, indicating a strong influence on magnetization dynamics by superconductivity. We note that the shift of $H_r$ below $T_C$ is clearly seen in a series of S/F/S Josephson junctions with different thickness of $Ni_{80}Fe_{20}$ layer ranging from 5 to 30 nm (see Fig. S1-S4 in Supplemental Information).

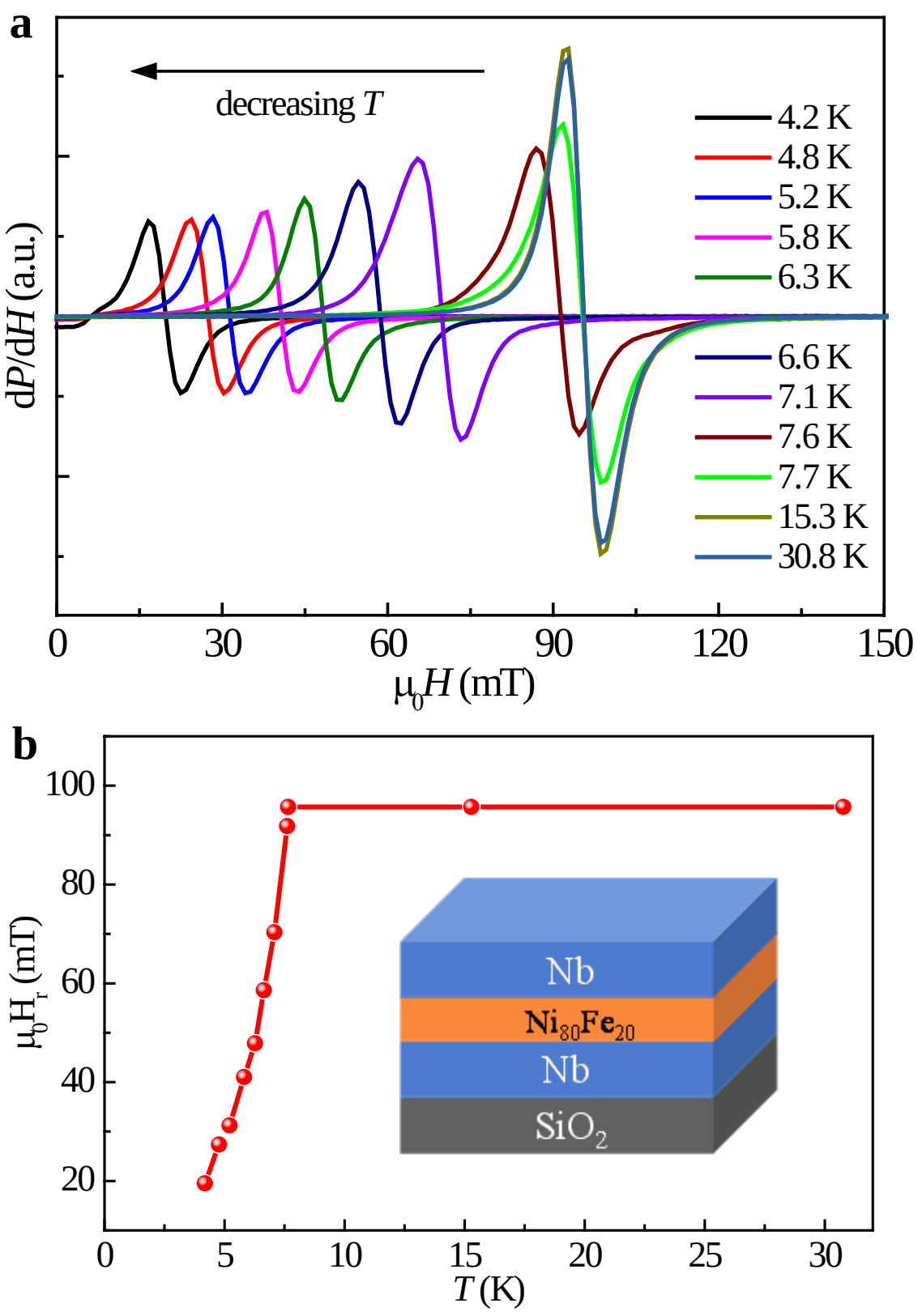

**Figure 2 FMR spectra of the Nb(100 nm)/$Ni_{80}Fe_{20}$(20 nm)/Nb(100 nm) Josephson junction. a**, The FMR spectra as a function of temperature. Below the superconducting transition temperature $T_C$ ~ 7.7 K, the resonance field $H_r$ shifts rapidly to a lower field with decreasing temperature. **b**, The resonance field $H_r$ as a function of temperature. The inset plots the structure of the sample. The significant shift of $H_r$ below $T_C$ evidences a strong influence on magnetization dynamics induced by superconductivity.

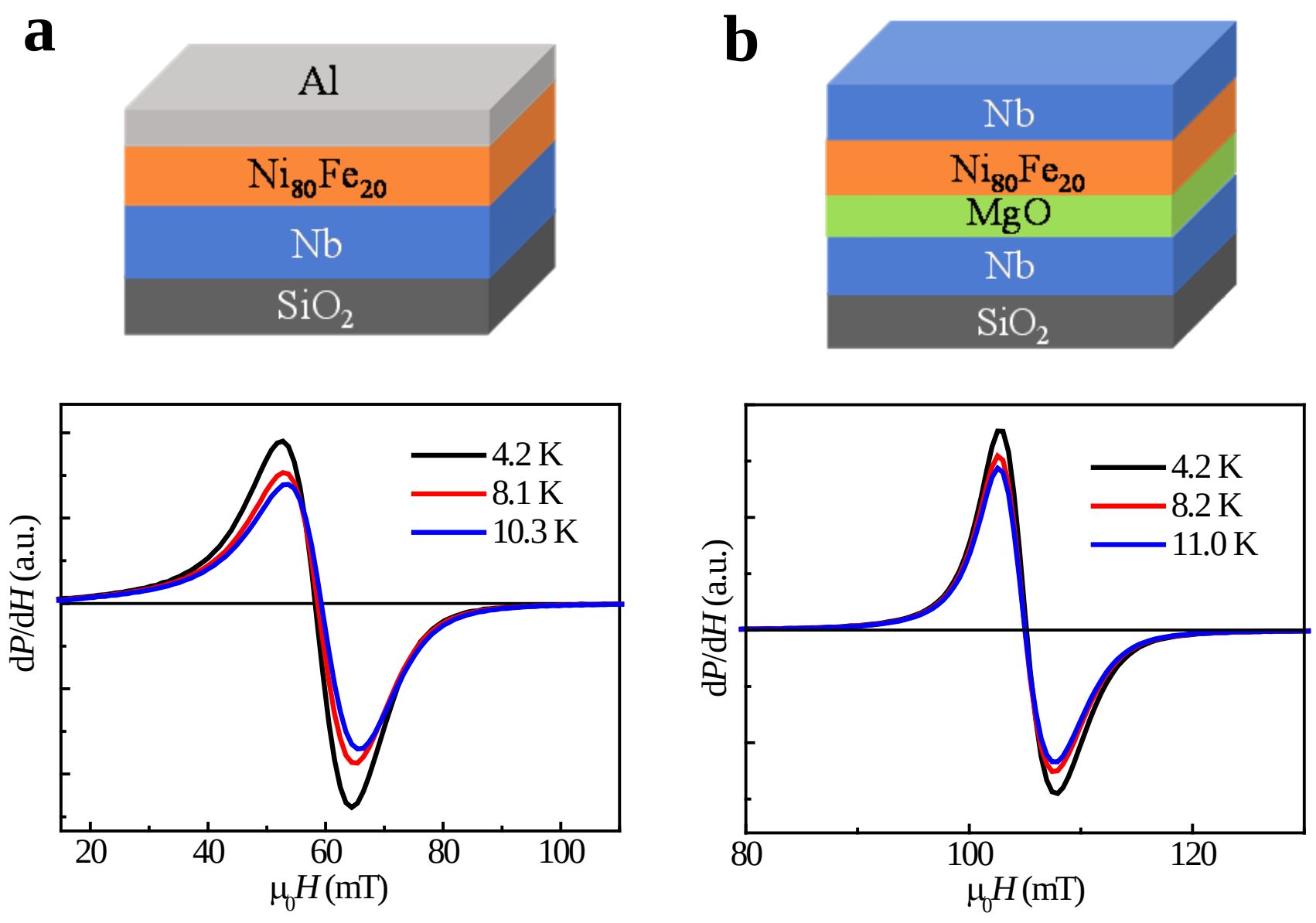

**Figure 3 Control experiments on S/F bilayer and S/I/F/S multilayer. a**, FMR spectra of a Nb(100 nm)/$Ni_{80}Fe_{20}$(20 nm) bilayer. The resonance field $H_r$ shifts little below the superconducting transition temperature $T_C$. **b**, FMR spectra of a Nb(100 nm)/MgO(10 nm)/$Ni_{80}Fe_{20}$(20 nm)/Nb(100 nm) multilayer. The resonance field $H_r$ does not shift below $T_C$. These control experiments suggest that the shift of $H_r$ is caused by supercurrents passing through the ferromagnetic layer rather than a local effect at one S/F interface.

For comparison, we also measured the FMR of a Nb(100 nm)/$Ni_{80}Fe_{20}$(20 nm) bilayer. As shown in Fig. 3a, for the S/F bilayer, $H_r$ does not shift obviously below $T_C$. From 10 K to 4.2 K $H_r$ only shifts about 1 mT. This observation is similar to a previous FMR study of S/F bilayers[22]. This control experiment clarifies that the shift of $H_r$ is closely related to the geometry of ferromagnetic Josephson junctions rather than one S/F interface. According to previous studies, the saturation magnetization $M_s$ of ferromagnetic layer changes little (< 1%) below $T_C$ by the interaction with superconductivity in the S/F/S trilayers and multilayers[32,33]. Thus, it can not account for the significant shift of $H_r$ (~ 70 mT) below $T_C$.

The shift of $H_r$ to a lower field indicates that an effective inner magnetic field parallel to the external magnetic field is produced in the superconducting state. In other words, there should be an extra torque induced by superconductivity to assist the external field torque to keep the magnetization precession. As this extra torque below $T_C$ is observed in S/F/S junctions but not in the S/F bilayer, it implies that supercurrents passing through Josephson junctions, rather than a local effect at one S/F interface, could play a critical role. To verify this viewpoint, we have performed another control experiment in a S/I/F/S multilayer where the supercurrents are blocked by the insulating MgO layer.

For typical S/MgO/S Josephson junctions, the thickness of MgO layer is usually below 2 nm. Above 2 nm the wave function can not overlap and tunneling supercurrents will be blocked. We then made a S/I/F/S multilayer with the thickness of insulating MgO layer of 10 nm. The FMR spectra of this insulating multilayer is presented in Fig. 3b. No obvious shift of $H_r$ is observed below $T_C$. From 11 K to 4.2 K, $H_r$ only shifts about 1 mT. This second control experiment further confirms that the extra torque below $T_C$ is due to supercurrents passing through the ferromagnetic layer. Since singlet supercurrents do not carry a net spin and should not cause a spin-transfer torque, it is concluded that the extra torque below $T_C$ is induced by triplet supercurrents.

In the following, we discuss the mechanism of spin-transfer torques induced by triplet supercurrents in the S/F/S Josephson junctions. The FMR experiments in our study is a situation of ferromagnetic Josephson junctions with precessing magnetization. Several theoretical models[28-31] have discussed on this situation and predicted that the long range triplet supercurrents can be stimulated by varying *in time* (rather than *in space*) the orientation of the magnetization in the ferromagnet. The microwave causes a precession of the magnetization, which corresponds to the pumping of a uniform ($q$ = 0) mode of magnons to the system. The presence of pumped magnons leads to the long-range triplet proximity effect. The pumped magnon spin current is compensated by the spin current carried by the triplet Cooper pairs due to the total spin angular momentum conservation. At the frequency of FMR, the critical supercurrent is greatly amplified due to the precessing magnetization and the generated triplet supercurrents through the junctions in turn exert a torque on the precessing magnetization.

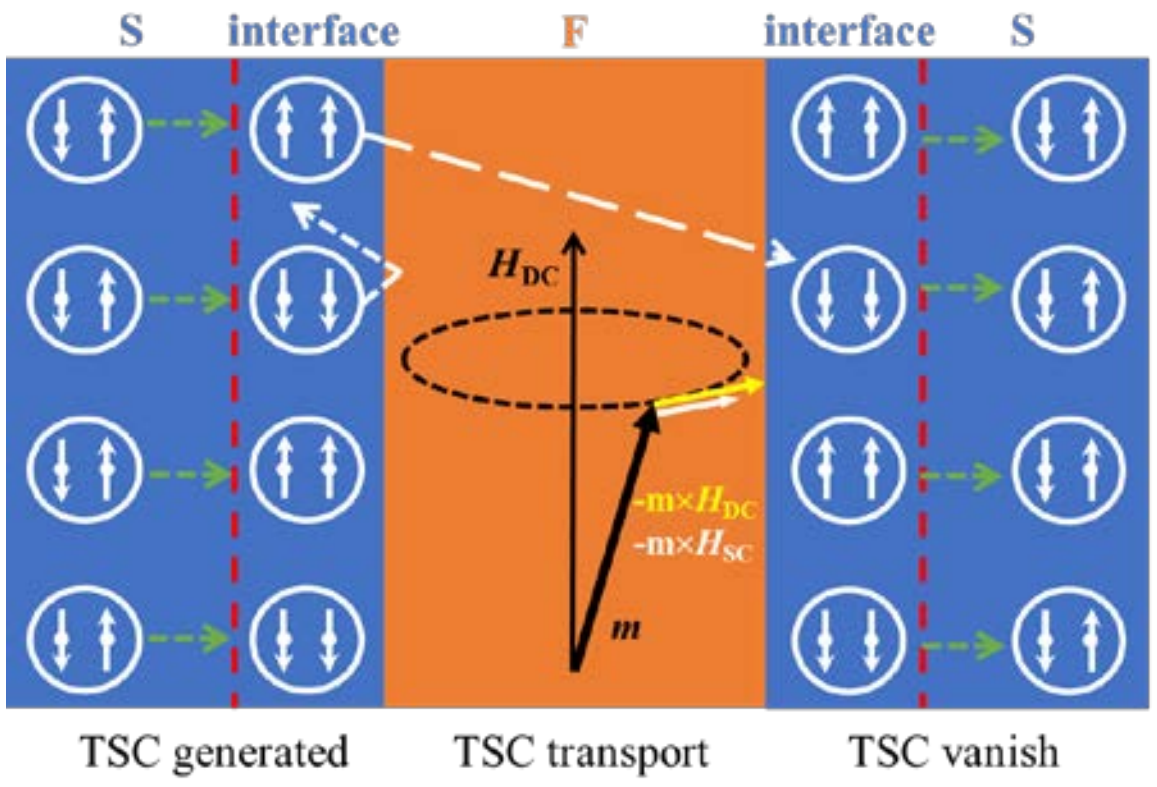

**Figure 4 Schematic illustration of triplet supercurrents (TSC) induced spin-transfer torque in S/F/S Josephson junctions with precessing magnetization.** Away from the S/F interfaces, only singlet Cooper pairs can exist. Triplet cooper pairs are generated at the interfaces due to the precessing magnetization. The triplet pairs with up spins (parallel to the external DC magnetic field) can transport through the F layer whereas the pairs with down spins are reflected back. The triplet pairs passing through the F layer exert a spin-transfer torque on the magnetization, causing a shift of resonance field at a fixed microwave frequency.

Figure 4 presents a schematic illustration of the dynamic process in the ferromagnetic junctions. Away from the S/F interfaces, only spin-singlet Cooper pairs can exist below $T_C$. A conversion from spin-singlet pairs to spin-triplet pairs due to the spin-mixing or spin-flip scattering occurs at the interfaces. The dynamically precessing magnetization plays an important role not only for the conversion process but also for the formation of Cooper pairs, by which the coherent charge and spin transport takes place through the junction due to the conservation of total spin angular momentum carried by triplet pairs and magnons[28-30]. For the triplet pairs with up spins (parallel to the external DC magnetic field), they can pass through the F layer. However, for the triplet pairs with down spins, they will be reflected back to the interface. Then triplet Cooper pairs passing through the ferromagnetic metal can exert a torque on the magnetization. This torque has the same direction as the torque generated by the DC magnetic field. As a consequence, the resonance field $H_r$ shifts to a lower field.

In summary, our FMR experiments in a series of S/F/S Josephson junctions demonstrate a significant modification on magnetization dynamics induced by superconductivity. In contrast, such a phenomenon is absent in S/F bilayers and S/I/F/S multilayers. Therefore, these results provide a strong evidence for the existence of spin-transfer torques induced by long-range triplet supercurrents. The observation of spin-transfer torque associated with triplet supercurrents as well as its strong influence on magnetization dynamics paves a way to the applications of superconducting spintronic devices.

## Methods

The superconductor-ferromagnet heterostructures including Nb/$Ni_{80}Fe_{20}$/Nb Josephson junctions and Nb/$Ni_{80}Fe_{20}$/Al(cap) bilayer are fabricated using dc magnetron sputtering on glass substrates. The base pressure of the sputtering system is about $10^{-6}$ Pa. The films are deposited at an Ar pressure of 0.5 Pa. The MgO layer in the Nb/MgO/$Ni_{80}Fe_{20}$/Nb multilayer is deposited by radio frequency sputtering with an Ar pressure about 0.8 Pa.

The FMR measurements are performed in a JEOL JA-200 spectrometer with an X-band ($f \approx 9.0$ GHz) cavity resonator. The system is equipped with a variable temperature unit down to liquid helium temperature.

## Acknowledgments

The authors are grateful to Prof. Jian-Wang Cai, Dr. Qin-Li Lu, and Mr. Yan Wen for help in sample preparation. This work was supported by the National Natural Science Foundation of China (Grant Nos. 11534015 and 51371192) and the National Key Research and Development Program of China (Grant No. 2016YFA0300700). Y.S. also acknowledges the support from Chinese Academy of Sciences (Grants No. XDB07030200 and KJZD-EW-M05).

## Author contributions

Y.S. and Y.Z. initialized this study. Y.Z. prepared several samples and carried out the transport measurements. L.L. performed FMR measurements. X.Z. contributed to materials and data analysis. Y.S. and Y.Z. wrote the paper, and all authors reviewed the paper.

# Supplemental Information

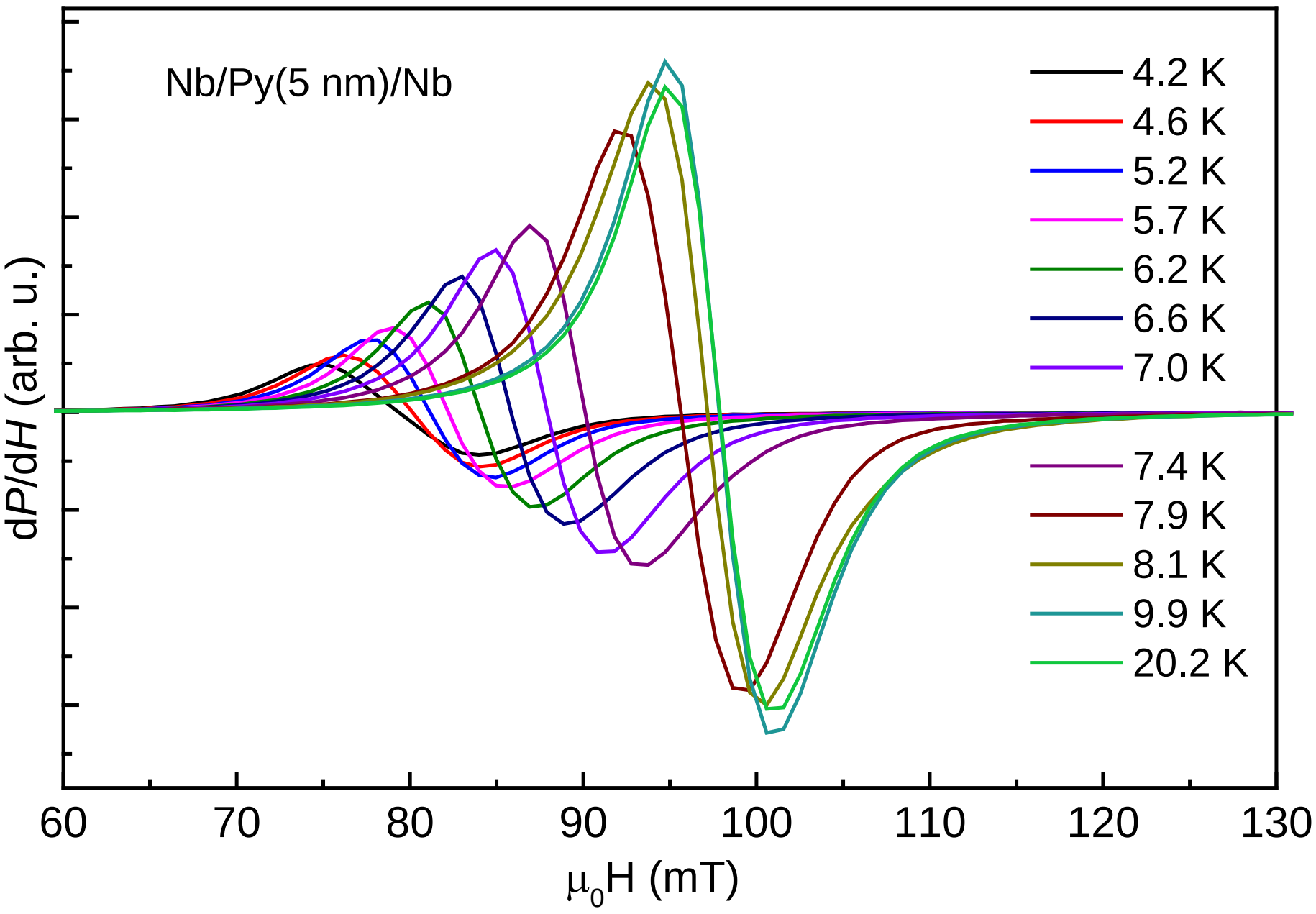

Figure S1 FMR spectra of a Nb(100 nm)/$Ni_{80}Fe_{20}$(5 nm)/Nb(100 nm) Josephson junction.

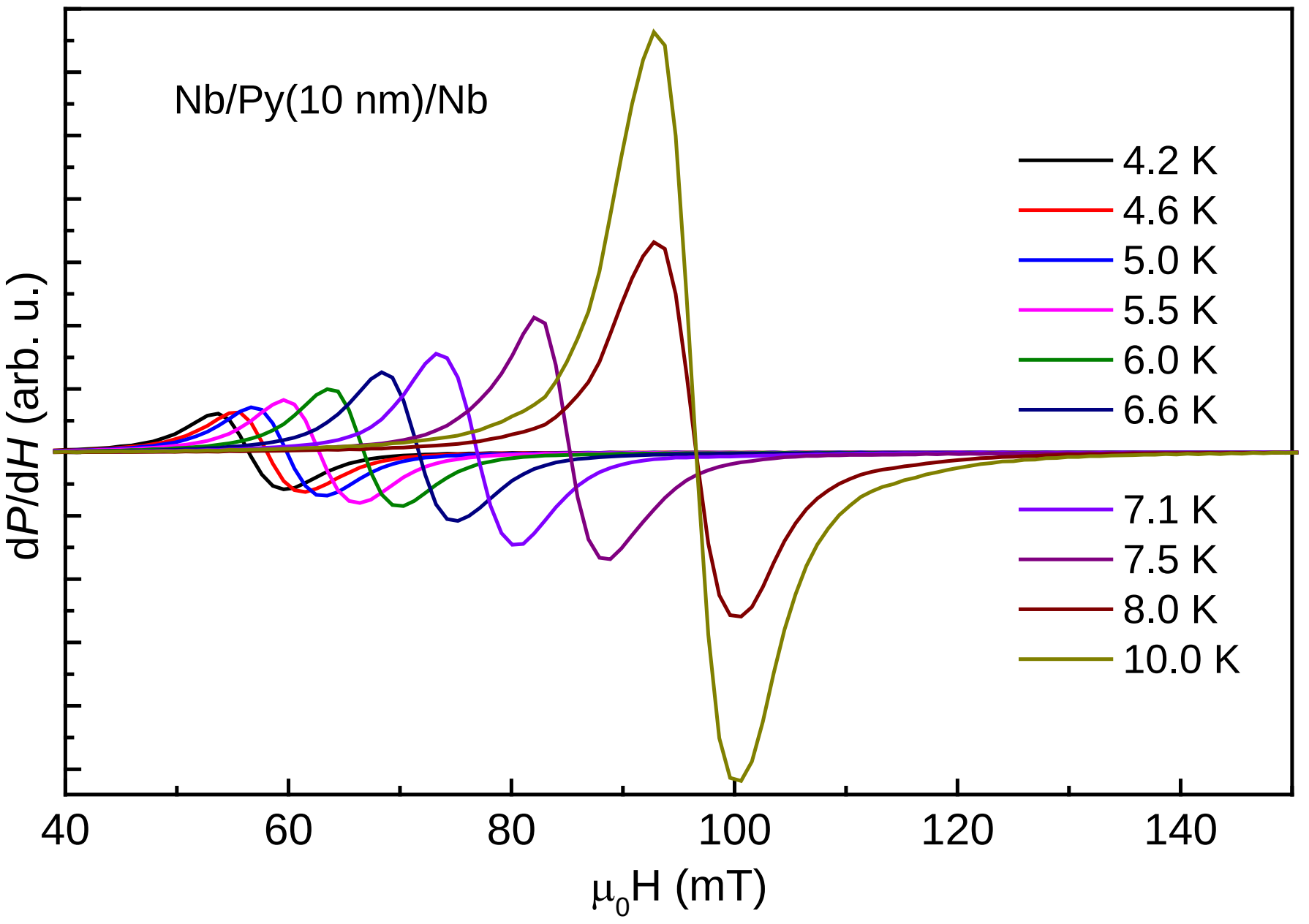

Figure S2. FMR spectra of a Nb(100 nm)/$Ni_{80}Fe_{20}$(10 nm)/Nb(100 nm) Josephson junction.

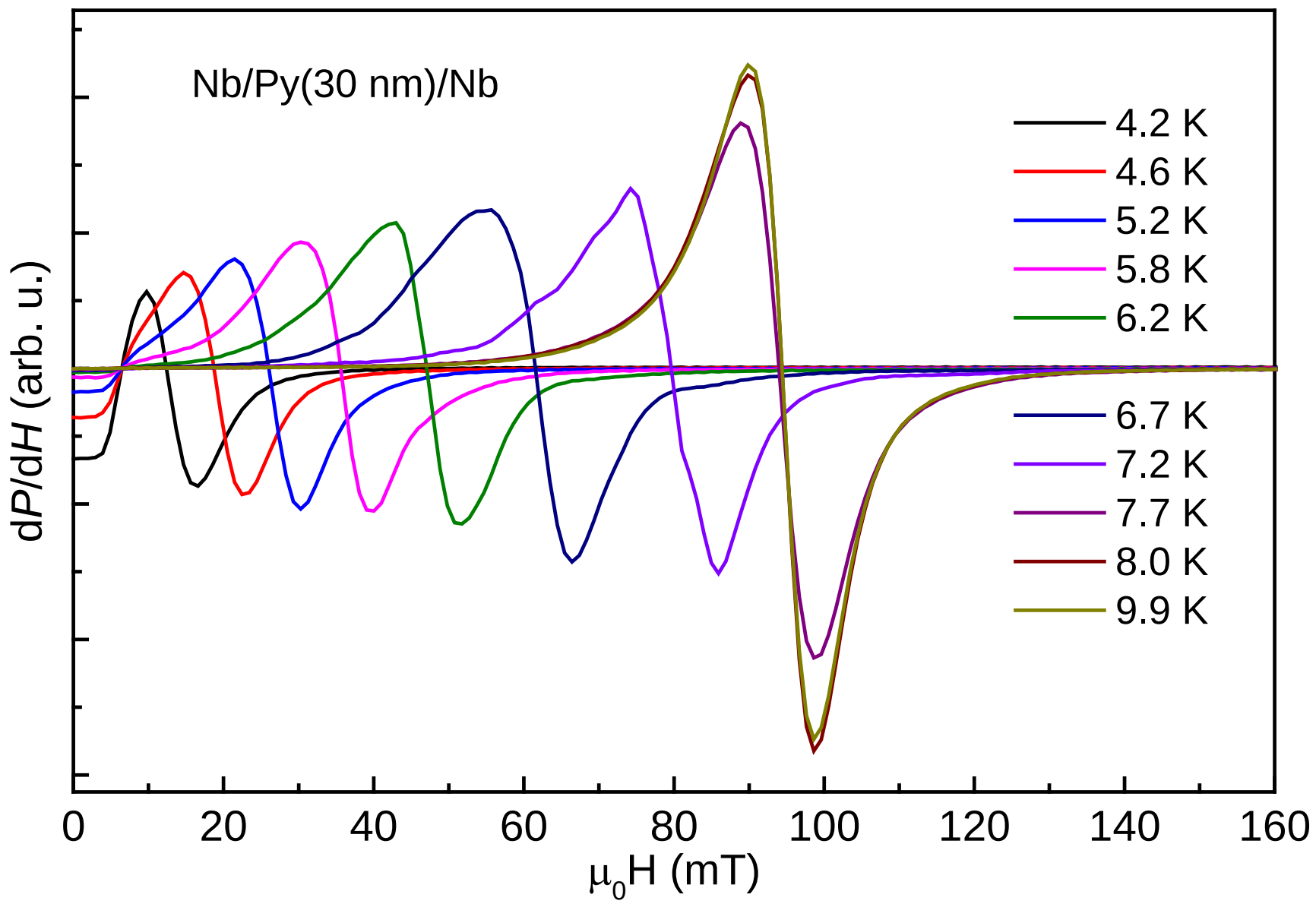

Figure S3. FMR spectra of a Nb(100 nm)/$Ni_{80}Fe_{20}$(30 nm)/Nb(100 nm) Josephson junction.

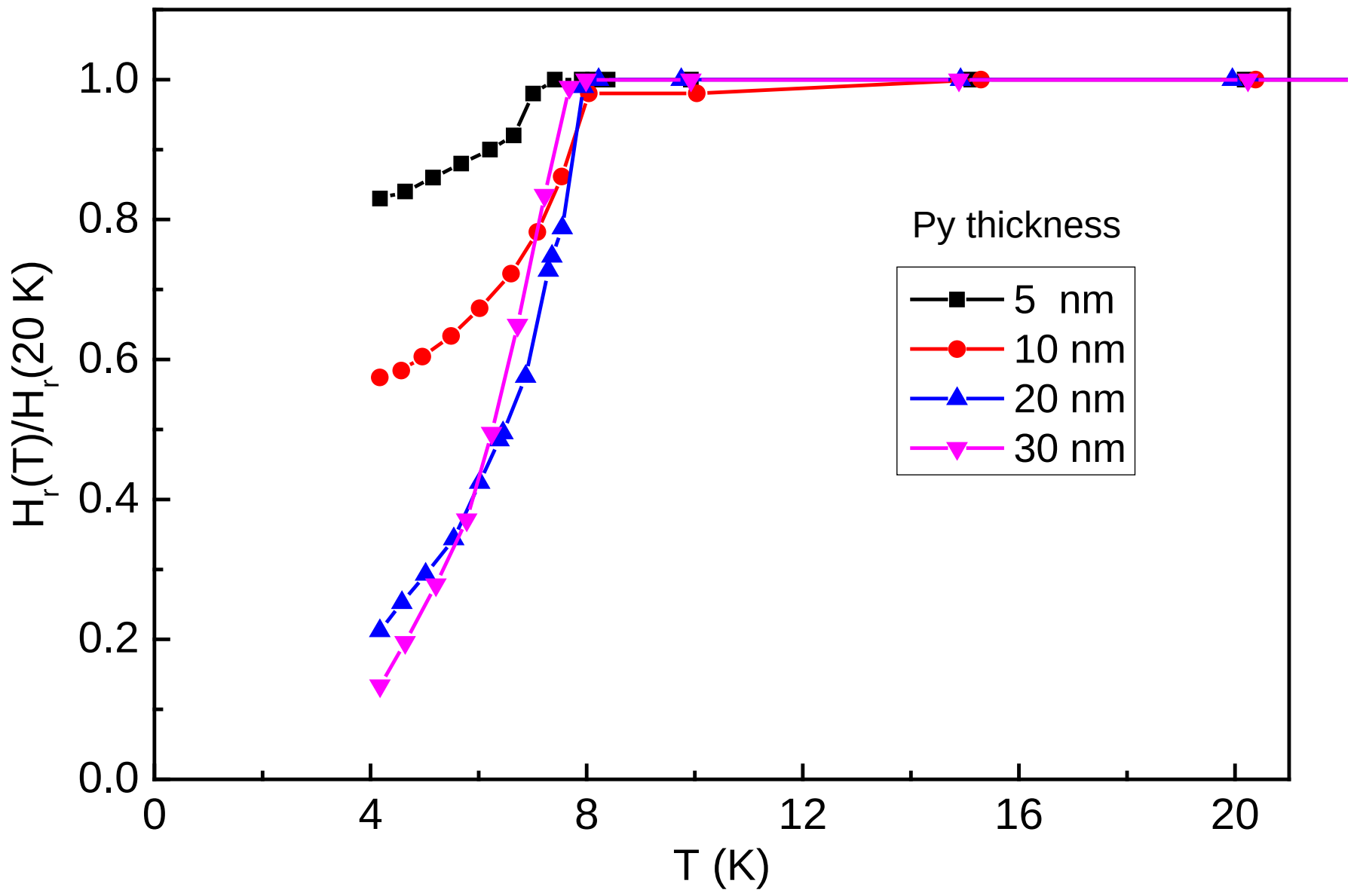

Figure S4. The relative shift of resonance field, $H_r$(T)/$H_r$(20 K), for a series of Nb/$Ni_{80}Fe_{20}$/Nb Josephson junctions with different thickness of $Ni_{80}Fe_{20}$ layer.